\documentclass[11pt]{article}
\begin{document}
\title
{Noncommutative inspired Schwarzschild black hole, Voros product and Komar energy}

\author{
{\bf {\normalsize Sunandan Gangopadhyay}$^{a,b}
$\thanks{sunandan.gangopadhyay@gmail.com, sunandan@iucaa.ernet.in, sunandan@bose.res.in}}\\
$^{a}$ {\normalsize Department of Physics, West Bengal State University, Barasat, India}\\
$^{b}${\normalsize Visiting Associate in Inter University Centre for Astronomy $\&$ Astrophysics,}\\
{\normalsize Pune, India}\\[0.3cm]
}
\date{}

\maketitle

\begin{abstract}
The importance of the Voros product in defining a noncommutative Schwarzschild black hole is shown. 
The entropy is then computed and the area law is shown to hold upto order $\frac{1}{\sqrt{\theta}}e^{-M^2/\theta}$. 
The leading correction to the entropy (computed in the tunneling formalism) is shown to be logarithmic.
The Komar energy $E$ for these black holes is then
obtained and a deformation from the conventional identity $E=2ST_H$ is found at the order $\sqrt{\theta}e^{-M^2/\theta}$.
This deformation leads to a nonvanishing Komar energy at the extremal point $T_{H}=0$ of these black holes.
Finally, the Smarr formula is worked out. Similar features also exist for a deSitter--Schwarzschild geometry\footnote{This presentation is based on the work in references \cite{sunplb, sungrg}.}.

\end{abstract}

\maketitle


\noindent In this presentation, we discuss some of the issues of noncommutative inspired
Schwarzschild black hole \cite{Smail, Smailrev} which has gained considerable interest recently. 
The primary point of interest is that there is no clear cut connection of this type of noncommutativity with 
standard notions of a noncommutative (NC) spacetime where point-wise multiplications are replaced by appropriate star multiplications. 
We shall point out that the Voros star product \cite{voros} plays an important role in
obtaining the mass density of a static, spherically symmetric, smeared, particle-like gravitational source. 
Our second objective is to derive quantum corrections to the semiclassical Hawking temperature and
entropy by the tunneling mechanism by going beyond the standard semiclassical approximation \cite{ban} for this black hole spacetime.
Finally, we would like to review the status of the relation between the Komar energy \cite{komar}, entropy and 
Hawking temperature ($E=2ST_H$) in the context of NC inspired Schwarzschild black holes.

The first issue can be addressed by taking recourse to the formulation and interpretational aspects of NC quantum
mechanics \cite{gouba, sunprl}. We observe that the inner product of the coherent
states $|z, \bar{z})$ (used in the construction of the wave-function of a “free point particle”) 
can be computed by using a deformed completeness relation (involving the Voros product) among the coherent states
\begin{eqnarray}
\int \frac{\theta dzd\bar{z}}{2\pi}~|z, \bar{z})\star(z, \bar{z}|=1_{Q}
\label{eg6}
\end{eqnarray}
where the Voros star product between two functions 
$f(z, \bar{z})$ and $g(z, \bar{z})$ is defined as
\begin{eqnarray}
f(z, \bar{z})\star g(z, \bar{z})=f(z, \bar{z})
e^{\stackrel{\leftarrow}{\partial_{\bar{z}}}
\stackrel{\rightarrow}{\partial_z}} g(z, \bar{z})~.
\label{eg7}
\end{eqnarray}
The wave-function of the “free point particle” on the NC plane \cite{spal, gouba} is given by
\begin{eqnarray}
\psi_{\vec{p}}=(p|z, \bar{z})=\frac{1}{\sqrt{2\pi\hbar^{2}}}
e^{-\frac{\theta}{4\hbar^{2}}\bar{p}p}
e^{i\sqrt{\frac{\theta}{2\hbar^{2}}}(p\bar{z}+\bar{p}z)}
\quad;\quad p=p_{x}+ip_{y}~,~ z=\frac{1}{\sqrt{2\theta}}(x+iy)
\label{wavefunction}
\end{eqnarray}
where the momentum eigenstates are normalised such that
$(p'|p)=\delta(p'-p)$ and satisfy the completeness relation
\begin{eqnarray}
\int d^{2}p~|p)(p|=1_{Q}~.
\label{mom_comp}
\end{eqnarray}
It turns out that a consistent probabilistic interpretation of this wave-function
can be given only when the Voros product is incorporated.
With these observations and interpretations in place,
we now write down the overlap of two coherent states 
$|\xi, \bar{\xi})$ and $|w, \bar{w})$ using the
completeness relation for the position eigenstates in eq.(\ref{eg6}) 
\begin{eqnarray}
(w, \bar{w}|\xi, \bar{\xi})=\int \frac{\theta dzd\bar{z}}{2\pi}
~(w, \bar{w}|z, \bar{z})\star(z, \bar{z}|\xi, \bar{\xi})~.
\label{overlap}
\end{eqnarray}
A simple inspection shows that 
$(w, \bar{w}|z, \bar{z})=\frac{1}{\theta}e^{-|\omega-z|^2}$ satisfies
the above equation. A straightforward dimensional lift of this solution
from two to three space dimensions immediately motivates one to write down the
mass density of a static, spherically symmetric,
smeared, particle-like gravitational source in three space dimensions as 
\begin{eqnarray}
\rho_{\theta}(r)=\frac{M}{(4\pi\theta)^{3/2}}
\exp\left(-\frac{r^2}{4\theta}\right)~.
\label{massden}
\end{eqnarray}  
The above discussion clearly brings out the important
role played by the Voros product in defining the mass density of the NC
Schwarzschild black hole.
Solving Einstein's equations with the above mass density incorporated
in the energy-momentum tensor leads to the following 
NC inspired Schwarzschild metric \cite{Smail},\cite{Smailrev}
\begin{eqnarray}
ds^2 = -\left(1-\frac{4M}{r\sqrt\pi}\gamma(\frac{3}{2},
\frac{r^2}{4\theta})\right)dt^2 + \left(1-\frac{4M}{r\sqrt\pi}
\gamma(\frac{3}{2},\frac{r^2}{4\theta})\right)^{-1} 
dr^2 + r^2(d\tilde\theta^2+\sin^2\tilde\theta d\phi^2)~. 
\label{1.04}
\end{eqnarray}
The event horizon of the black hole can be 
found by setting $g_{tt}(r_h)=0$ in eq.(\ref{1.04}), which yields
\begin{eqnarray}
r_h=\frac{4M}{\sqrt\pi}\gamma(\frac{3}{2},\frac{r^2_h}{4\theta}).
\label{1.05}
\end{eqnarray}
The large radius regime ($\frac{r_{h}^2}{4\theta}>>1$) is taken
where one can expand the incomplete gamma function to 
solve $r_h$ by iteration. Keeping upto next to leading order 
$\sqrt{\theta}e^{-{M^2}/{\theta}}$, we find
\begin{eqnarray}
r_h \simeq 2M\left[1-\frac{2M}{\sqrt{\pi\theta}}
\left(1+\frac{\theta}{2M^{2}}\right)e^{{-M^2}/{\theta}}
\right]~.  
\label{1.06}
\end{eqnarray}
Now for a general stationary, static and spherically 
symmetric space time, the Hawking temperature ($T_H$) 
is related to the surface gravity ($\kappa$) 
by the following relation \cite{Hawking}
\begin{eqnarray}
T_H=\frac{\kappa}{2\pi} 
\label{1.061}
\end{eqnarray}
where the surface gravity of the black hole is given by
\begin{eqnarray}
\kappa = \frac{1}{2}\left[\frac{dg_{tt}}{dr}\right]_{r=r_{h}}.
\label{1.07}
\end{eqnarray}
Hence the Hawking temperature for the NC inspired Schwarzschild black hole upto order 
$\sqrt{\theta}e^{-{M^2}/{\theta}}$ is given by
\begin{eqnarray}
T_{H}=\frac{1}{8{\pi}M}
\left[1-\frac{4M^3}{{\theta}{\sqrt{\pi\theta}}}
\left(1-\frac{\theta}{2M^{2}}-\frac{\theta^2}{4M^{4}}\right)
{e^{-M^2/\theta}}\right]~.
\label{1.08}
\end{eqnarray}
The first law of black hole thermodynamics can now be used to work out the Bekenstein-Hawking entropy. 
The law reads  
\begin{eqnarray}
dS_{BH}=\frac{dM}{T_H}~.
\label{1.1}
\end{eqnarray}
Hence the Bekenstein-Hawking entropy in the next to 
leading order in $\theta$ is found to be 
\begin{eqnarray}
S_{BH}=\int{\frac{dM}{T_H}}=4\pi M^2-16\sqrt{\frac{\pi}{\theta}}M^3
\left(1+\frac{\theta}{M^2}\right)e^{-M^2/\theta}~.
\label{1.11}
\end{eqnarray}
To express the entropy in terms of the NC horizon area ($A_{\theta}$), we use eq.(\ref{1.06}) to get
\begin{eqnarray}
A_{\theta}&=& 4\pi r^2_h=16\pi M^2-64\sqrt{\frac{\pi}{\theta}}
M^3\left(1+\frac{\theta}{2M^2}\right)e^{-M^{2}/\theta}+\mathcal{O}(\theta^{3/2}e^{-M^{2}/\theta}).
\label{1.12}
\end{eqnarray}
Comparing equations (\ref{1.11}) and (\ref{1.12}), 
we find that at the leading order in $\theta$ (i.e. upto order $\frac{1}{\sqrt{\theta}}e^{-{M^2}/{\theta}}$), 
the NC black hole entropy satisfies the area law 
(in the regime $\frac{r^2_h}{4\theta}>>1$)
\begin{eqnarray}
S_{BH}=\frac{A_{\theta}}{4}~.
\label{1.13}
\end{eqnarray}
We now look for corrections to the semiclassical area law upto leading order in $\theta$.

\noindent To do so, we first compute the corrected Hawking temperature $\tilde{T}_{H}$. For that we use the tunneling method by going beyond the semiclassical approximation \cite{ban}. Considering the massless scalar particle tunneling
under the background metric (\ref{1.04}),  the corrected Hawking temperature is given by
\begin{eqnarray}
\tilde{T}_{H}=T_{H}\left[1+\sum_{i}\frac{\tilde{\beta}_{i}\hbar^{i}}
{(Mr_{h})^{i}}\right]^{-1}~.
\label{corr_temp}
\end{eqnarray}
Applying the first law of black hole thermodynamics once again with this corrected Hawking temperature, gives the following expression for the corrected entropy/area law :
\begin{eqnarray}
S&=& \frac{A_{\theta}}{4\hbar}+2\pi\tilde{\beta}_{1}\ln A_{\theta} - \frac{64\pi^{2}\tilde{\beta}_{2}\hbar^2}{A_{\theta}}+\mathcal{O}(\sqrt{\theta}e^{-\frac{M^2}{\theta}})~\nonumber\\        &=& S_{BH}+2\pi\tilde{\beta}_{1}\ln S_{BH}-\frac{16\pi^{2}\tilde{\beta}_{2}\hbar}{S_{BH}}+\mathcal{O}(\sqrt{\theta}e^{-\frac{M^2}{\theta}})~.
\label{corr_entr}
\end{eqnarray}
Finally, we proceed to investigate the status of the relation between the Komar energy $E$,
entropy $S$ and Hawking temperature $T_H$
\begin{eqnarray}
E=2ST_{H}
\label{komar_ener}
\end{eqnarray}
in the case of these NC inspired black holes. The expression for the Komar
energy $E$ for the NC inspired Schwarzschild metric (\ref{1.04}) is given by \cite{sunplb}
\begin{eqnarray}
E=\frac{2M}{\sqrt\pi}\gamma\left(\frac{3}{2},\frac{r^2}{4\theta}\right)
-\frac{Mr^3}{2\theta\sqrt{\pi\theta}}e^{-r^2/(4\theta)}~.
\label{komarmass}
\end{eqnarray}
We therefore identify $M$ as the mass of the black hole since $E=M$ in the limit $r\rightarrow\infty$. This identification
plays an important role as we shall see below.

\noindent The above expression computed near the event horizon of the black hole\footnote{The Komar energy computed near the event horizon of the black hole plays an important role in obtaining the coefficient of the logarithmic
correction term (\ref{corr_entr}) in the entropy \cite{sunplb}.}
upto order $\sqrt{\theta}e^{-{M^2}/{\theta}}$ gives
\begin{eqnarray}
E &=& M\left[1-\frac{2M}{\sqrt{\pi\theta}}
\left(\frac{2M^{2}}{\theta}+1\right)e^{{-M^2}/{\theta}}-\frac{1}{M}\sqrt{\frac{\theta}{\pi}}e^{{-M^2}/{\theta}}\right].
\label{komar1c}
\end{eqnarray}
Finally, using eqs.(\ref{1.08}), (\ref{1.11}) and (\ref{komar1c}),       
we obtain
\begin{eqnarray}
E&=&2ST_{H} +2\sqrt{\frac{\theta}{\pi}}e^{{-M^2}/{\theta}}+ 
\mathcal{O}(\theta^{3/2}e^{-M^{2}/\theta})\nonumber\\
&=&2ST_{H} +2\sqrt{\frac{\theta}{\pi}}e^{{-S}/{(4\pi\theta)}}+ 
\mathcal{O}(\theta^{3/2}e^{-S/(4\pi\theta)})
\label{komar2a}
\end{eqnarray}
where in the second line we have used eq.(\ref{1.11}) to replace $M^2$
by $S/(4\pi)$ in the exponent. Interestingly, we find that the
relation $E=2ST_H$ gets deformed upto order $\sqrt{\theta}e^{-{M^2}/{\theta}}$ which is consistent with 
the fact that the area law also gets modified at this order. The deformation also yields a nonvanishing Komar energy at the extremal point $T_H =0$ of these black holes \cite{sungrg}. Also, we have once again managed to write down the deformed relation in terms of the Komar energy $E$, entropy $S$ and the Hawking temperature $T_H$. Similar features are also present for a de-Sitter Schwarzschild geometry \cite{dym}.
Eq.(\ref{komar2a}) can also be written with $M$ being expressed in terms of the black hole parameters
$S$ and $T_H$ using eq.(\ref{komar1c}) 
\begin{eqnarray}
M&=&2ST_{H} +\frac{1}{2\pi\sqrt{\pi\theta}}\left(S+\frac{S^2}{2\pi\theta}+6\pi\theta\right)e^{-S/(4\pi\theta)}
+\mathcal{O}(\theta^{3/2}e^{-S/(4\pi\theta)}).
\label{nc_smarr}
\end{eqnarray} 
We name eq.(\ref{nc_smarr}) as the {\it{Smarr formula}} \cite{smarr} {\it{for NC inspired Schwarzschild black hole}}
since $M$ has been identified earlier to be the mass of the black hole. 


\end{document}